# Novel magnetic ordering in LiYbO$_2$ probed by muon spin relaxation

Eric M. Kenney[1], Mitchell M. Bordelon[2], Chennan Wang[4], Hubertus Luetkens[4], Stephen D. Wilson[3], and Michael J. Graf [1]

[1] Department of Physics, Boston College, Chestnut Hill, Massachusetts, 02467, USA

[2] Materials Physics and Applications–Quantum, Los Alamos National Laboratory, Los Alamos, New Mexico, 87545, USA

[3] Materials Department, University of California, Santa Barbara, California, 93106, USA

[4] Laboratory for Muon Spin Spectroscopy, Paul Scherrer Institute, 5232 Villigen, Switzerland

## Abstract

The stretched diamond lattice material LiYbO$_2$ has recently been reported to exhibit two magnetic transitions ($T_{N1}$ = 1.1 K, $T_{N2}$ = 0.45 K) via specific heat, magnetization, and neutron scattering measurements [Bordelon et al., Phys. Rev. B **103**, 014420 (2021)]. Here we report complementary magnetic measurements down to $T$ = 0.28 K via the local probe technique of muon spin relaxation. While we observe a rapid increase in the zero-field muon depolarization rate at $T_{N1}$, for $T < T_{N1}$ we do not observe the spontaneous muon precession which is typically associated with long-range magnetic ordering. The depolarization rate in the ordered state shows a surprising sensitivity to magnetic fields applied along the initial spin polarization direction. Using a simple one-dimensional model, we show that these results are consistent with the unusual random-phase bipartite incommensurate magnetic structure proposed by Bordelon et al. for the intermediate temperature range $T_{N2} < T < T_{N1}$. We also find evidence for magnetic fluctuations persisting to our lowest temperatures, but no obvious signature of the transition or spontaneous muon precession at and below $T_{N2}$, respectively.

While a variety of interesting phenomena are expected to occur in magnetically frustrated systems, there are relatively few material families that predictably host frustrated magnetism and typically systems order or freeze at finite temperatures [1–4]. There are even fewer frustrated three-dimensional (3D) systems in the low-spin $S = \frac{1}{2}$ limit. 3D frustration has mainly focused on magnetic pyrochlore lattices, such as the lanthanide materials $Ln_2M_2O_7$, with $Ln$ =Lanthanide and $M$ = Metal or Metalloid [5]. Magnetic frustration on diamond lattices has been relatively less researched.

Recently, Bordelon et al. [6] reported the successful synthesis of polycrystalline $LiYbO_2$, along with a detailed set of structural, specific heat, magnetization, and neutron scattering data. Structurally the system forms a stretched diamond lattice. The $Yb^{3+}$ ions can be modeled as two interpenetrating face-centered cubic sublattices with a nearest-neighbor Heisenberg interaction $J_1$ between sublattices and a next-nearest-neighbor interaction $J_2$ within a sublattice. The $Yb^{3+}$ ions have electronic moments with $J_{\text{eff}} = \frac{1}{2}$ and $\mu_{eff} = 2.74\,\mu_B$. Above $1.1$ K the system is paramagnetic, with a Curie-Weiss temperature of $\Theta_{\text{CW}} = -3.4$ K. Specific heat showed a transition at 1.1 K, followed by a much weaker second transition at 0.45 K, resulting in a modest frustration factor of $f \approx 3$. Neutron diffraction measurements show that below 450 mK each of the two $Yb^{3+}$ sublattices order as an incommensurate spiral with a propagation vector $K = (0.384, \pm0.384, 0)$, and a rotational phase difference of $0.58\,\pi$ between the sublattices; a phase difference close to $\pi$ is expected as the nearest neighbor exchange $J_1$ (which is between sublattices) is antiferromagnetic. This magnetic structure was shown to be well described within the framework of a Heisenberg $J_1 - J_2$ Hamiltonian as applied to the stretched diamond lattice, with $\frac{J_2}{|J_1|} \approx 1.42$.

The intermediate state between 450 mK and 1.1 K is less well understood. Unusual neutron diffraction patterns in this temperature interval were best described by an effective model with two helical structures on the sublattices with the same $K$ as the low-temperature phase but allowing for random variations of the phase difference between the two $Yb^{3+}$ sublattices, for example different phase differences in different magnetic domains. Averaging over sublattices with different relative phases was then used to model the neutron data This novel approach suggests that the $Yb^{3+}$ moments first order within their respective sublattices at 1.1 K, and then the two sublattices lock into the aforementioned $0.58\,\pi$ phase difference below 450 mK. To our knowledge, this is the first time a two-step magnetic transition of this type has been proposed. The simple Heisenberg model cannot account for this intermediate temperature phase, and its appearance was attributed to additional anisotropic exchange terms in the Hamiltonian. Finally, inelastic neutron scattering (INS) revealed the existence of low-energy fluctuations of about 1 meV down to 38 mK. These low-energy fluctuations persisted in applied fields up to 10 T, at which a field-polarized state was induced. However, it was not clear from these results if the fluctuations were caused by conventional magnetic excitations within the ordered state or by more exotic mechanisms, e.g. fluctuations between degenerate ground states or within a 'quantum spiral spin liquid' state [7-9].

In this work we present positive muon spin relaxation ($\mu^+$SR) measurements as a local-probe complement to the bulk measurements reported in Ref. 6. Our results confirm that LiYbO$_2$ magnetically orders at $T_{N1} \sim 1.1$ K, but the spontaneous oscillatory depolarization in zero applied field that is typical for long-range magnetic order is absent. We present a simple model for muon depolarization spectra for incommensurate bipartite lattices and show that the random phase model proposed for the LiYbO$_2$ magnetic structure would naturally suppress coherent muon spin oscillations, consistent with our observations. Moreover, we find that the model predicts an unusual sensitivity of the muon depolarization to weak applied longitudinal fields despite the strong depolarization in zero field. Finally, we find persistent magnetic fluctuations down to T = 0.28 K, but cannot resolve a second magnetic transition near 0.45 K.

A polycrystalline sample of LiYbO$_2$ was prepared using a solid-state reaction between Yb$_2$O$_3$ and Li$_2$CO$_3$ as reported previously [6]. Sample purity was verified via x-ray diffraction and susceptibility measurements. The powder was pressed into a disk approximately 1 cm in diameter and 3 mm thick in an Ar-atomosphere glovebox, and minimal exposure to air was maintained at all times.

The $\mu^+$SR experiments were performed at the Paul Scherrer Institute using the General Purpose Surface-Muon (GPS) [10] and Dolly instruments on the $\pi$M3 and $\pi$E1 beamlines, respectively. Measurements in GPS were made using a gas flow cryostat between 40 and 1.5 K. Measurement in Dolly were made using a He-3 cryostat between 1.8 and 0.28 K; additionally, measurements were taken at 40 K in Dolly to extract background parameters by comparison with the GPS data. The samples were mounted on 25-$\mu$m thick copper foil in order to enhance thermalization at low temperatures. Measurements were performed in longitudinal spin-polarization mode with the initial muon polarization anti-parallel to the beam momentum. Data were analyzed using the MUSRFIT program [11].

The time-dependent muon depolarization between 40 K and 0.28 K are shown in Fig 1. Above 10 K the depolarization is temperature independent, and dominated by the Li nuclei. It is well described by

$$A(t) = A_0 \big[ (1 - F_B) \, G_{KT}(t) e^{-\lambda t} + F_B e^{-\lambda_B t} \big] \ . \quad (1)$$

The function $G_{KT}(t)$ is the Gaussian Kubo-Toyabe [12–14] given by

$$G_{KT}(t) = \frac{1}{3} + \frac{2}{3} (1 - \sigma_N^2 t^2) e^{-\frac{\sigma_N^2 t^2}{2}} \ . \quad (2)$$

The first term in Eq. (1) describes muons depolarizing in the sample, while the second term accounts for a fraction $F_B$ of muons landing in the cryostat and sample holder. For data taken in GPS, $F_B$ is negligibly small, while $F_B = 0.15(3)$ and $\lambda_B = 0.27 \ \mu s^{-1}$ are obtained in Dolly, with $F_B$ determined as described above. We find $\sigma_N = 0.163(1) \ \mu s^{-1}$, typical for compounds containing lithium, which has a fairly large nuclear moment of 3.3 $\mu_N$. An additional exponential depolarization is present in the sample, with $\lambda = 0.17(3) \ \mu s^{-1}$ in this temperature range, presumably due to fluctuating $Yb^{3+}$ moments. In the crossover region between 10 and 2 K, the $Yb^{3+}$ moments begin to slow and thus dominate the local field and muon depolarization. We note this temperature range corresponds to the broad maximum in specific heat originating from the onset of electronic correlations [6].

We now focus on our primary region of interest $T < 2$ K. In Fig. 2 we show representative short-time depolarization curves. An abrupt change occurs at 1.1 K, signaling the onset of magnetic order. The data is noteworthy for the lack of spontaneous muon precession as typically observed in materials with long-range magnetic order. In the inset we show a high-resolution curve taken at short times at $T = 0.28$ K, and the lack of any oscillations is clear. The data below 2 K are well described by the phenomenological function

$$A(t) = A_0 \left\{ 1 - F_B \right\} \left( (1 - f_\lambda) e^{-\frac{(\sigma t)^2}{2}} + f_\lambda e^{-\lambda t} \right) + F_B e^{-\lambda_B t} \right\}, \quad (3)$$

as shown by the solid lines in Figs. 1 and 2. In Fig. 3 we show the parameters $\sigma$, $\lambda$, and $f_\lambda$ as a function of temperature. The abrupt change in all three parameters at 1.10 K is clear. A fraction $(1 - f_\lambda)$ of muons are rapidly depolarizing with an initially positive curvature approximated by a Gaussian decay, characteristic of a quasi-static array of densely packed moments. At the lowest temperatures $\sigma$ is roughly 45 MHz, indicating a characteristic internal field of order $\sigma/\gamma_\mu = 530$ G; $\gamma_\mu = 0.08514 \, MHz/G$ is the muon gyromagnetic ratio. A fraction $f_\lambda = \frac{1}{3}$ (termed the 'tail') is expected and observed below 1 K, representing the ensemble-averaged fraction of muons lying parallel to the local magnetic field, with decay caused by magnetic fluctuations and/or dilute magnetic impurities in the sample. Surprisingly, none of the parameters clearly indicate any transition at T = 0.45 K where a weak anomaly is observed in specific heat [6].

In Fig. 4, we show results for the depolarization at $T = 0.28$ K for several magnetic fields applied along the initial muon spin polarization direction (longitudinal field, or 'LF'). Depolarization is suppressed when the LF is comparable to or greater than the internal field experienced by the muon, resulting in an increase in $f_\lambda$. A small field of 50 G immediately suppresses some of the long-time depolarization, due in part to the weakly magnetic background contribution. However, the detailed LF data is not well-described by the 'standard' model [12], and we will elaborate on this in the discussion below. We also note that a slow relaxation of the tail remains even at higher fields, due to depolarization in the sample by magnetic fluctuations. Such fluctuations were observed via neutron scattering in Ref. 6. At LF values of several hundred Gauss, the tail then lifts further, as expected for the internal field of order 530 G as inferred from our fit results at low temperatures.

We now discuss our $\mu^+$SR results in the context of the observations reported in Ref. 6. Our results are clearly consistent with the onset of a phase transition at 1.1 K, and the sharpness of this transition (see Fig. 3) suggests that there is very little chemical or magnetic disorder. Nonetheless, no spontaneous muon precession is observed. In fact, many systems with complex ordering such as multiple-Q [13] or incommensurate spiral [14] phases exhibit clear magnetic order based on neutron scattering results, but with no muon precession. This is typically presented as a qualitative result of broadened field distribution at the muon site without supporting calculations. Here, the specific magnetic structure proposed for $T_{N2} < T < T_{N1}$ based on the neutron scattering results – the random-phase bipartite incommensurate (RPBI) state - allows for an *analytical calculation* of the muon depolarization. We approximate the internal field distribution as two identical, but independent, incommensurate internal field distributions. The internal field distribution seen by the muon ensemble, $D_{RPBI}(B_{loc})$, is then described by the convolution of these two distributions:

$$D_{\text{RPBI}}(B_{\text{loc}}) = (D_{\text{inc}} * D_{\text{inc}})(B_{\text{loc}}) \qquad (4)$$

where $D_{\text{inc}}(B_{\text{loc}})$ is the basic model for the field distribution seen by muons inside an incommensurate magnet [15]:

$$D_{\text{inc}}(B_{\text{loc}}; B_{\text{max}}) = \begin{cases} \frac{1}{\pi} \frac{1}{\sqrt{B_{\text{max}}^2 - B_{\text{loc}}^2}}, & -B_{\text{max}} < B_{\text{loc}} < B_{\text{max}} \\ 0, & \text{Otherwise} \end{cases} \qquad (5)$$

Equation (4) can be solved analytically. The result is a complete elliptical integral of the first kind, $K[x]$, and

$$D_{\text{RPBI}}(B_{loc}; B_{\text{max}}) = \frac{4}{\pi^2} K\left[1 - \left(\frac{B_{loc}}{2B_{\text{max}}}\right)^2\right] \{-2B_{\text{max}} \le B_{\text{loc}} \le 2B_{\text{max}}\} \qquad (6)$$

We plot the field distributions for $D_{\text{inc}}$ and $D_{\text{RPBI}}$ in Figure 5. Notably, the field distribution resulting from this convolution is strongly peaked about zero field, as the random phases will result in a significant contribution to the field distribution of $B_{loc}$ where $+B_{max}$ from one spiral will be negated by the $-B_{max}$ from the other spiral. Qualitatively, the random phases are a form of magnetic disorder, peaked at $B_{loc} = 0$ which dominates the shape of the internal field distribution. It can be shown (see Supplemental Material) that the resulting depolarization function is

$$P(t) = \frac{1}{3} + \frac{2}{3} J_0^2 (\gamma_\mu B_{\text{max}} t) \qquad (7)$$

where $J_0$ is the zeroth-order Bessel function of the first kind, $B_{\text{max}}$ is the asymptotically maximum field produced by a single magnetic sublattice and $\gamma_\mu$ is the muon gyromagnetic ratio. The square of the Bessel function is in contrast to the usual first-power Bessel oscillation found for the distribution given in Eq. 5.

In the case of slow spin dynamics, such as the fluctuations observed in INS at low temperatures, static polarization functions generalize by multiplying the static tail by an exponential:

$$P(t) = \frac{1}{3} e^{-\lambda t} + \frac{2}{3} J_0^2 (\gamma_\mu B_{\text{max}} t). \qquad (8)$$

For experimental data with a background, the asymmetry function then becomes

$$A(t) = A_0 \left[ \{1 - F_B\} \left( (1 - f_\lambda) J_0^2 + f_\lambda e^{-\lambda t} \right) + F_B e^{-\lambda_B t} \right]. \qquad (9)$$

where $f_\lambda \cong \frac{1}{3}$. This corresponds to our phenomenological function in Eq. (3), but with the Gaussian factor replaced by $J_0^2$.

In Fig. 5 we compare the resulting depolarization functions to the data at 550 mK along with phenomenological Gaussian fits described above by placing the fit parameters from extracted equation (3) into (9) The resulting oscillations in the RPBI model are smaller than the spread of our data, even without considering other potential sources of disorder which would further suppress oscillations, such as lattice defects. Finally, we note that both the Gaussian and Bessel-squared functions have the same short-time limit

$$e^{-\frac{(\sigma t)^2}{2}} \approx J_0^2(\sigma t) \approx 1 - \frac{1}{2}\sigma^2 t^2 \quad (10)$$

which would explain the good agreement of our phenomenological fit function Eq. 3 to the data.

The unusual shape of $D_{\mathrm{RPBI}}(B_{\mathrm{loc}}; B_{\mathrm{max}})$ will also affect the depolarization in applied longitudinal fields. In general the application of a field $B_{\mathrm{LF}}$ comparable to the internal field $B_{\mathrm{int}}$ will increase the non-oscillatory fraction $f_\lambda$ and when $B_{\mathrm{LF}} >> B_{\mathrm{int}}$ one finds $f_\lambda \sim 1$. This field dependence has been calculated for materials with simple commensurate order (oscillating as a cosine function) and colinear incommensurate order (oscillating as a Bessel function) [12, 15]. We expect that for the RPBI model the system will be very sensitive to the application of a LF due to the peak at zero in the field distribution. We have calculated the field dependence of $f_\lambda$ for the RPBI model, and the result is shown in Fig. 6a, along with calculations for commensurately and colinear incommensurately ordered materials. We see that for increasingly complex ordered materials the low-field response becomes stronger for a given maximum field. While we do not have data for the LF response in the RPBI state, for comparison we have also included in the figure the extracted field-dependent value for $f_\lambda$ as determined by fitting data taken at T = 0.28 K (fixed-phase regime). The experimental response is qualitatively the same as that calculated using the RPBI model, although being significantly lower in magnitude. Intriguingly, scaling the applied $B_{LF}$ by a factor of $\alpha = 1/3$ gives near-perfect match between the data and the RPBI calculation (Fig. 6b). This observation, together with the nearly identical zero-field muon response for $T_{N1} < T < T_{N2}$ and $T < T_{N2}$, suggests that the underlying physical processes governing muon depolarization are nearly the same in both the intermediate and low-temperature phases.

The low temperature fixed-phase state is itself complex, with two doubly degenerate spiral incommensurate orders, and so we expect depolarization in this phase to be similar to that in the RPBI phase. Moreover, neutron scattering shows that fluctuations are present throughout the temperature range studied here. Combined, these two factors may produce very similar muon depolarization in zero and longitudinal fields. Presuming this is the case, the good agreement in LF when including a scaling factor for the applied field is reminiscent of behavior observed in other systems. $Yb_2Ti_2O_7$ is a geometrically frustrated system with persistent spin fluctuations deep in the ordered phase and a complex magnetic order [16]. It exhibits an unusual form of secondary dynamics known as 'sporadic dynamics,' which results in linear screening of applied longitudinal fields as developed to describe fluctuations in the frustrated Kagome material $SrCr_8Ga_4O_{19}$ [17]. The sporadic model treats the local field at the stopping site as being intermittent, having a non-zero value for only a fraction $\alpha$ of the time. This results in a linear reduction of $B_{LF}$ due to simple scaling arguments. Additionally, the shape of the depolarization curve is unaffected by the sporadic dynamics, causing to the system to appear static or quasi-static despite the presence of fluctuations. For $Yb_2Ti_2O_7$, the screening factor is $\alpha \cong 0.17$ [18]. At present it is unclear, however, if the observed scaling behavior in $LiYbO_2$ results from sporadic dynamics, or is simply a consequence of the different levels of complexity between the RPBI and fixed-phase magnetic structures. Further detailed studies, preferably on single crystals, are required to better understand these fluctuations within the two magnetic phases.

   Summarizing, we performed local probe $\mu^+SR$ measurements on $LiYbO_2$ to complement recent neutron, bulk magnetic, and thermodynamic measurements [6]. We find clear signatures of the sharp magnetic transition at 1.1 K, but none for the second observed transition at 0.45 K. No spontaneous muon precession is observed for $T$ down to 0.28 K, and we propose a simple model based on the novel magnetic structure reported in Ref. [6] that is consistent with this result. This model is also consistent with our observation of unusual sensitivity of the muon depolarization to applied longitudinal magnetic fields. Finally, our results confirm the presence of magnetic fluctuations down to $T = 0.28$ K, suggesting that despite the relatively low frustration factor of $f \sim 3$, in $LiYbO_2$ fluctuations amongst allowed spiral states/configurations may remain significant. Future studies on single crystals to lower temperatures will help to clarify many of the unanswered questions regarding this material.

## Acknowledgements

This work is based on experiments performed at the Swiss Muon Source SµS, Paul Scherrer Institute, Villigen, Switzerland. MMB and SDW were supported by the US Department of Energy Office of Basic Energy Sciences, Division of Materials Science and Engineering under award DE-SC0017752.

**Figures**

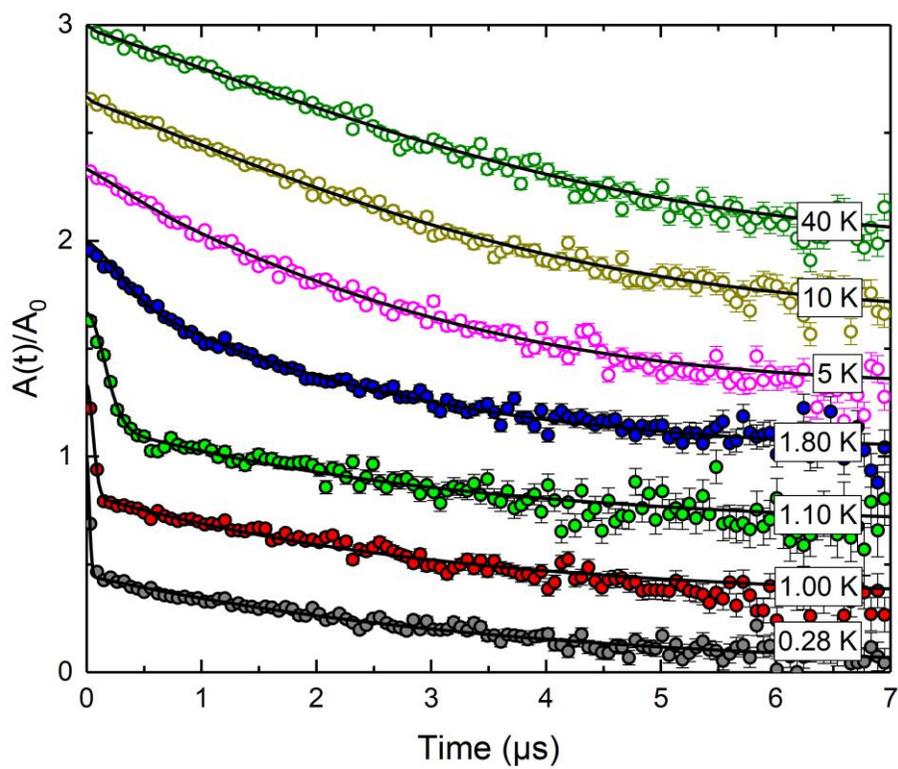

Figure 1: Zero field depolarization spectra of LiYbO₂ measured on the Dolly (solid circles) and GPS (empty circles) spectrometers with fits as described in text (solid lines). Curves are offset by equal amounts clarity.

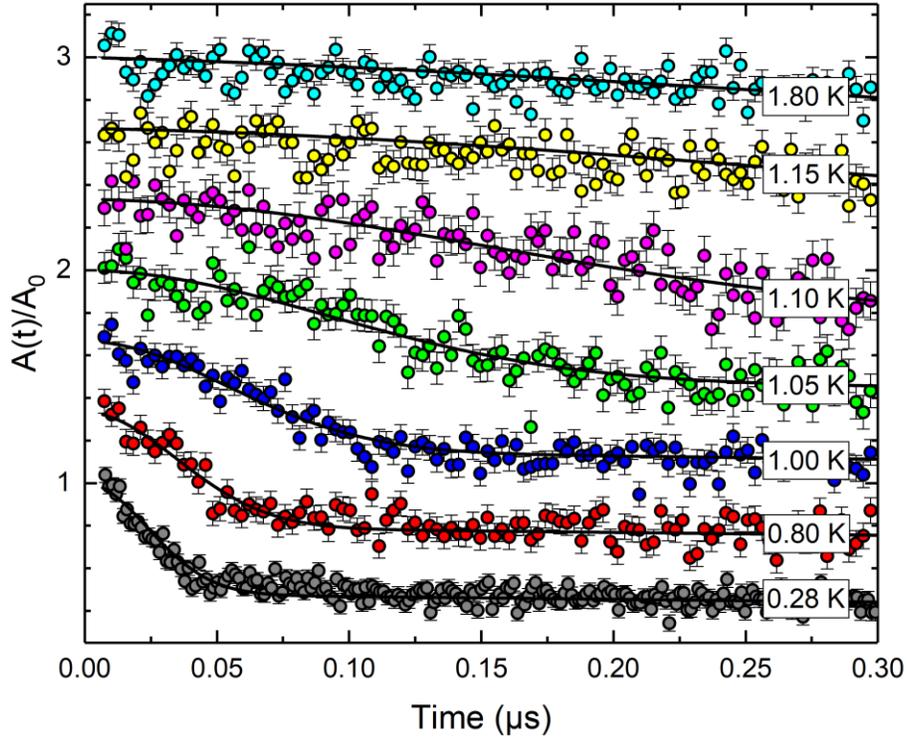

Figure 2: Short-time zero-field (ZF) asymmetry spectra plotted from 0 μs to 0.3 μs. Spectra were measured on Dolly and range from 0.28 K to 1.8 K. We observe a sharp transition between 1.10K and 1.15K. At 0.28 K the time-gating was adjusted to increase temporal resolution. No oscillations or dips are seen in the spectra, despite improved resolution. Curves are offset by equal amounts for clarity.

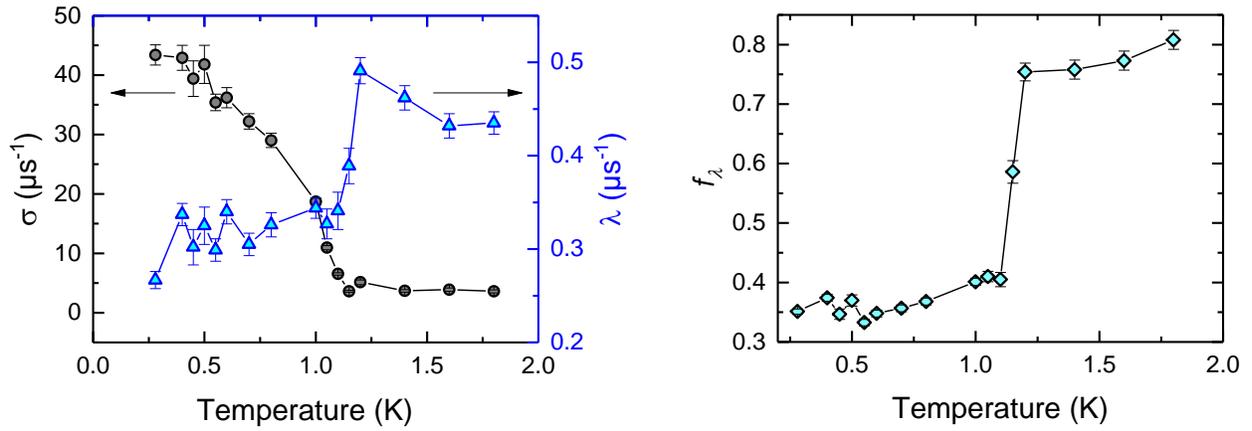

Figure 3: Temperature dependence of the fit parameters described in the main text. The exponential rate $\lambda$ and its fraction $f_\lambda$ describes the long-time decay of the asymmetry, presumably due to quasi-static fluctuations, while the Gaussian rate $\sigma$ describes the short-time behavior due to static order. The drop of $f_\lambda$ to $1/3$ below the transition is consistent with a long-range ordering transition reducing dynamics, resulting in the appearance of a weakly damped $1/3$-tail.

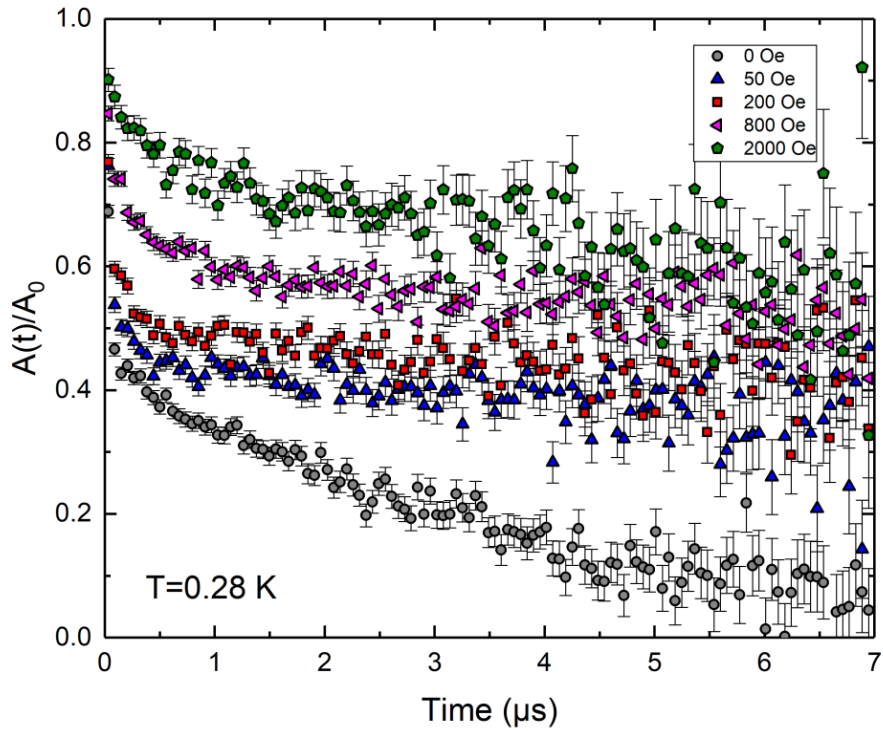

Figure 4: Longitudinal field μ⁺SR spectra at 0.28K. The initial lifting of the asymmetry spectra corresponds to the suppression of static disorder or quasi-static fluctuations. The small lifting of the asymmetry between 50 Oe to 800 Oe allows us to estimate the internal field strength associated with the rapid decay to be roughly 500 Oe.

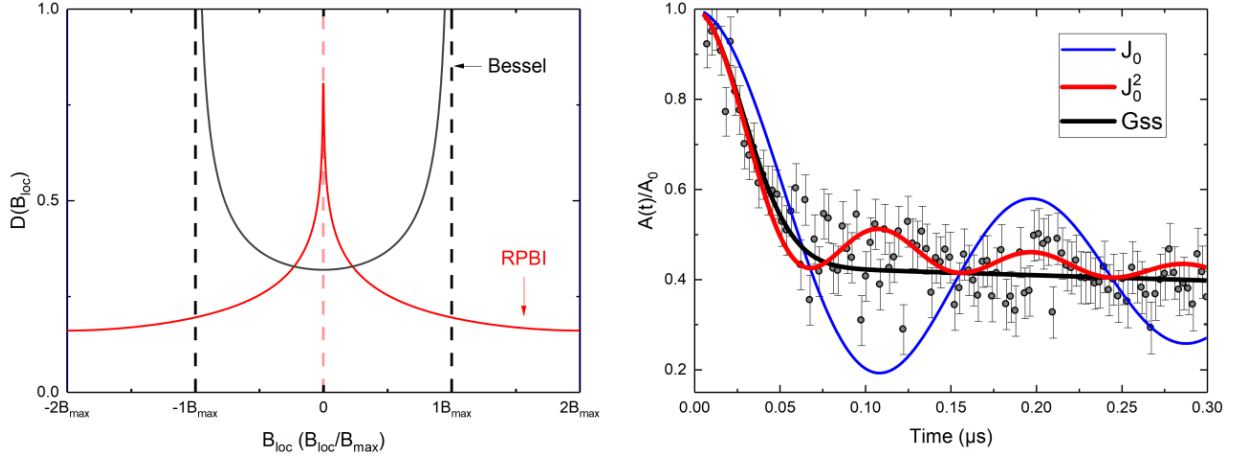

Figure 5: (Left) Field distributions $D(B_{loc})$ for a simple incommensurate magnet (Bessel) and the RPBI distribution, as described in the text. (Right) The temperature dependent asymmetry data (circles) and plots of the phenomenological fit (Gss), the Bessel function, and Bessel squared depolarization as described in the text. The parameters used are obtained from the shown Gaussian fit. For the Bessel depolarization, we take $\gamma_\mu B_{max} = \sqrt{2}\,\sigma$ in accordance with the short-time expansions of $J_0$ and $e^{\frac{-\sigma^2 t^2}{2}}$.

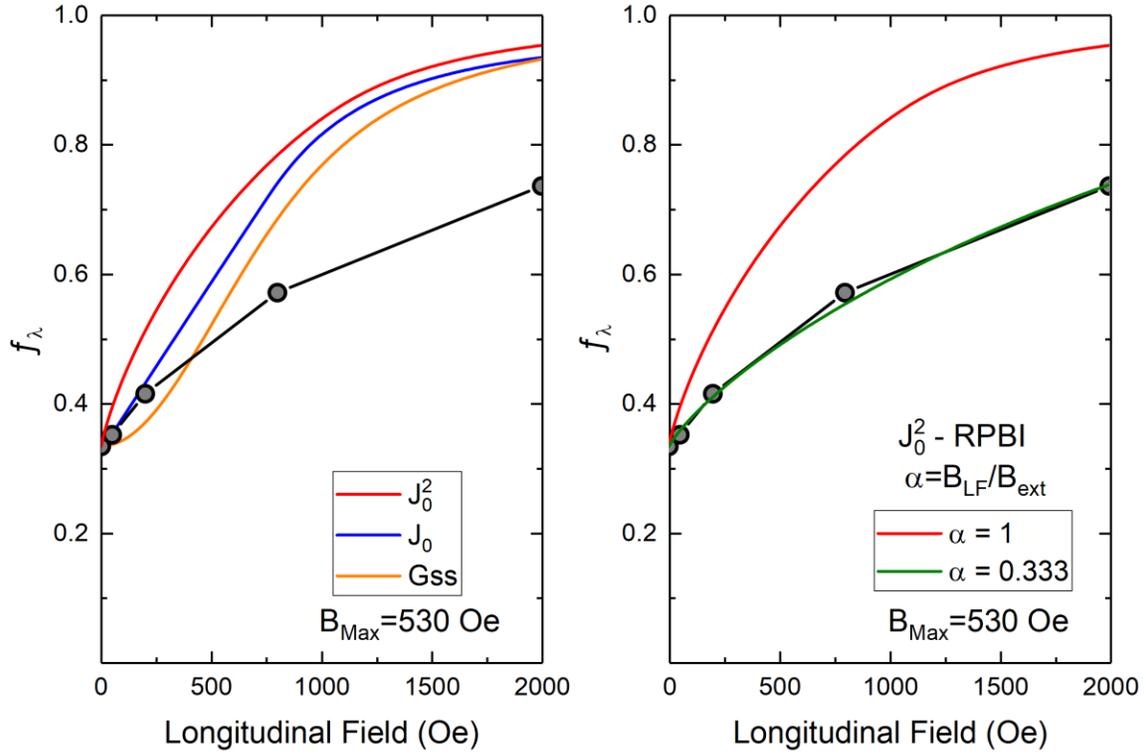

Figure 6: (Left) Longitudinal field dependence of the static tails for the data shown in Figure 4 (grey points with grey lines to guide the eye.) Overlaid is the calculated LF dependence for the field distributions in the text, given the fit parameters extracted from the fit at 0.28 K. The used internal field values are derived from the fitted depolarization rate. The lack of high-field agreement in any model indicates the presence of dynamics, despite the gaussian-like line-shape in zero-field. (Right) The LF tail recalculated using a linearly screened field with $\alpha = 1/3$.